\documentclass{ws-procs9x6}

\begin{document}

\title{LORENTZ-VIOLATING PHOTONS WITH A MASS TERM}

\author{MAURO CAMBIASO}

\address{Departamento de Ciencias F\'\i sicas, Universidad Andr\'es Bello\\
%Avenida Republica 220, 
Santiago, Chile\\
E-mail: mcambiaso@unab.cl}

\author{RALF LEHNERT}

\address{Indiana University Center for Spacetime Symmetries\\
Bloomington, IN 47405, USA\\
E-mail: ralehner@indiana.edu}

\author{ROBERTUS POTTING}

\address{CENTRA, Departamento de F\'\i sica, Universidade do Algarve\\ 8005-139 Faro, Portugal\\ 
E-mail: rpotting@ualg.pt}

\begin{abstract}
Perturbative calculations in quantum field theory 
often require the regularization of infrared divergences.
In quantum electrodynamics, 
such a regularization can for example be accomplished by 
a photon mass 
introduced via the Stueckelberg method. 
The present work extends this method
to the QED limit of the Lorentz- and CPT-violating Standard-Model Extension.
\end{abstract}

\bodymatter

\section{Introduction}
\label{intro}

Massive vector particles are of relevance for various subfields of physics 
including the weak interaction, 
test models for photon-mass searches,
and the regularization of certain infrared divergences 
in quantum-field calculations. 
In this latter context, 
the mass term should be introduced such that
the symmetries of the original model remain unspoiled.
For a U(1) Lorentz-invariant gauge theory,
this can be accomplished with the Stueckelberg method.\cite{Stueckelberg}

In recent years, 
Lorentz-violating quantum field theories 
have become a focus of theoretical\cite{SMEtheo} and experimental\cite{SMEexp} inquiry. 
Such theories also require the regularization of infrared divergences 
in some circumstances, 
and the question arises, 
as to whether the Stueckelberg method 
can be adapted to such Lorentz-breaking U(1) gauge theories.
As in the Lorentz-invariant case, 
the adapted Stueckelberg method should be compatible 
with all symmetries of the theory to be regulated,
so that the usual Stueckelberg approach can be employed
for the Lorentz-violating extension of conventional QED.
However,
the Lorentz violation in this model 
allows a broader range of compatible Stueckelberg terms 
that may be introduced, 
and this freedom can then be used 
to streamline calculations
in Lorentz-breaking QED.

The present work 
provides one possible class of extensions 
of the usual Stueckelberg procedure
to Lorentz-violating QED.
Section \ref{lagr} gives a brief overview 
of the procedure at the lagrangian level.
Some implications of the Lorentz-breaking Stueckelberg
Lagrangian are discussed in Sec.~\ref{EoM}.

\section{Lagrangian Analysis}
\label{lagr}

We begin with the usual 
free-photon Lagrangian
in the minimal Standard-Model Extension
coupled to a conserved source $j^\mu$:
\begin{eqnarray}
\label{lagrangian}
{\cal L}_{\gamma} \!& = &\!
-\tfrac{1}{4}F^2
-A\cdot j
-\tfrac{1}{4}(k_F)^{\kappa\lambda\mu\nu}
F_{\kappa\lambda}F_{\mu\nu}
+\tfrac{1}{2}\epsilon^{\kappa\lambda\mu\nu}(k_{AF})_\kappa A_\lambda F_{\mu\nu}\,.
\end{eqnarray}
Here,
Lorentz and CPT breakdown is controlled by 
the spacetime-constant backgrounds $k_F$ and $k_{AF}$. 

The absence of Lorentz symmetry in the above Lagrangian~(\ref{lagrangian}) 
allows us to drop the requirement of Lorentz invariance 
for the mass-type term $\delta {\cal L}_{m}$ to be introduced for the photon.
However, 
the inclusion of arbitrary Lorentz violation into  $\delta {\cal L}_{m}$ 
may be problematic:
consider the case in which $k_F$ and $k_{AF}$
are such that a subgroup of the Lorentz group remains unbroken.
A regulator violating this residual symmetry 
may be undesirable, 
so that the breakdown of the remaining invariant subgroup
may have to be excluded from $\delta {\cal L}_{m}$. 

For our present purposes, 
however,
we consider arbitrary Lorentz violation in $\delta {\cal L}_{m}$.
This yields more general results 
also relevant for purposes other than 
infrared regularization.
More specifically,
we implement the Stueckelberg procedure by
introducing a scalar field $\phi$ as follows:\cite{Cam12}
\begin{equation}
\label{Stueckelberg}
\delta {\cal L}_{m}=
\tfrac{1}{2}(\partial_\mu \phi-m A_\mu)
\hat{\eta}^{\mu\nu}(\partial_\nu \phi-m A_\nu)\,,
\end{equation}
where $m$ can later be identified with the photon mass
and
\begin{equation}
\label{hat-eta}
\hat{\eta}^{\mu\nu}=\eta^{\mu\nu}+\hat{G}^{\mu\nu}
\end{equation}
with $\hat{G}^{\mu\nu}$ a Lorentz-violating operator that 
may contain derivatives but is otherwise spacetime constant.
The small number of Lorentz-symmetric pieces 
still contained in an arbitrary $\hat{G}^{\mu\nu}$ 
can be removed if necessary.

As in the conventional Stueckelberg case,
the key feature of ${\cal L}_\gamma+\delta {\cal L}_{m}$ is 
its invariance (up to total derivatives)
under a local gauge transformation
\begin{equation}
\label{eq:gaugetransf}
\delta A_\mu = \partial_\mu \epsilon(x)\,,\quad 
\delta \phi = m \epsilon(x)\,.
\end{equation}
With the addition of $\xi$-type gauge fixing
$\mathcal{L}_\textrm{g.f.}=-\frac{1}{2\xi} (\partial_\mu \hat{\eta}^{\mu \nu}A_\nu + \xi m \phi)^2$ and a Faddeev--Popov contribution
$\mathcal{L}_\textrm{F.P.}$,
the model Lagragian 
${\cal L}={\cal L}_\gamma+\delta {\cal L}_{m}+\mathcal{L}_\textrm{g.f.}+ \mathcal{L}_\textrm{F.P.}$
becomes\cite{Cam12}
\begin{eqnarray}
\label{eq:totalL}
{}\hspace{-10mm}\mathcal{L}\! &=&\!-\tfrac{1}{4}F^2
-A\cdot j +\tfrac{1}{2}m^2 A_\mu \hat{\eta}^{\mu \nu} A_\nu
-\frac{1}{2\xi}(\partial_\mu \hat{\eta}^{\mu \nu}A_\nu)^2\nonumber\\
&&\! {}-\tfrac{1}{2}\phi(\partial_\mu\hat{\eta}^{\mu\nu}\partial_\nu+\xi m^2)\phi
- \bar c(\partial_\mu\hat{\eta}^{\mu\nu}\partial_\nu+\xi m^2)c\frac{{}}{{}}\nonumber\\
&&\!{}-\tfrac{1}{4}(k_F)^{\kappa\lambda\mu\nu}
F_{\kappa\lambda}F_{\mu\nu}
+\tfrac{1}{2}\epsilon^{\kappa\lambda\mu\nu}(k_{AF})_\kappa A_\lambda F_{\mu\nu}\,.
\end{eqnarray}
Note that 
the scalar $\phi$ and the ghosts $c$ and $\bar c$ are now decoupled
and can be integrated out.
We can therefore disregard these fields in what follows.

\section{Equations of Motion}
\label{EoM}

The Lagrangian~(\ref{eq:totalL}) yields the following 
equation of motion for the photon:
\begin{eqnarray}
\label{EoM1}
\left[\eta^{\mu\alpha}\eta^{\nu\beta}\partial_\mu
+(k_{AF})_\mu\epsilon^{\mu\nu\alpha\beta}
+(k_F)^{\mu\nu\alpha\beta}\partial_\mu\right]F_{\alpha\beta}&&\nonumber\\
{}+\left[m^2\hat{\eta}^{\mu\nu}
+\frac{1}{\xi}\hat{\eta}^{\mu\alpha}\hat{\eta}^{\nu\beta}\partial_\alpha\partial_\beta\right]A_\mu &=& j^\nu
\,.
\end{eqnarray}
Taking the 4-divergence of this equation yields
\begin{equation}
\label{longitudinal}
\left(\hat{\eta}^{\mu\nu}\partial_\mu\partial_\nu+\xi m^2
\right)(\partial_\alpha\hat{\eta}^{\alpha\beta}A_\beta)=0\,.
\end{equation}
Note that 
$(\partial_\alpha\hat{\eta}^{\alpha\beta}A_\beta)$ projects out 
one degree of freedom contained in $A_\mu$. 
Equation~(\ref{longitudinal}) establishes that 
this degree of freedom is not excited by the source $j_\nu$, 
so  $(\partial_\alpha\hat{\eta}^{\alpha\beta}A_\beta)$
is an auxiliary mode.
This is consistent with the expectation of three physical degrees of freedom 
for a massive vector field.

A plane-wave ansatz in Eq.~(\ref{EoM1}) yields the model's
dispersion relation; it has the following structure:\cite{Cam12}
\begin{equation}
\label{fullDR}
\frac{1}{\xi}
\big(1+\tfrac{1}{4}\hat{G}^\alpha_\alpha\big)
(\hat{\eta}^{\mu\nu}p_\mu p_\nu-\xi m^2)Q(p)=0\,.
\end{equation}
The first factor does not contain physical modes.
The $(\hat{\eta}^{\mu\nu}p_\mu p_\nu-\xi m^2)$ piece
corresponds to the auxiliary mode 
governed by Eq.~(\ref{longitudinal}).
The 
%$\xi$-independent 
factor $Q(p)$ 
is associated with the physical degrees of freedom
and has the structure
\begin{equation}
\label{physDR}
Q=(p^2-m^2)^3+
r_2(p^2-m^2)^2+
r_1(p^2-m^2)+
r_0\,.
\end{equation}
Here, 
the coefficients $r_j$ are coordinate scalars
containing the Lorentz-breaking tensors $k_F$ and $k_{AF}$.
They vanish in the limit $k_F, k_{AF}\to0$.
The explicit expressions for the $r_j$ 
can be found in the literature.\cite{Cam12}
Note that the physical dispersion relation~(\ref{physDR}) 
is consistent with the expectation of three conventional massive modes
perturbed by small Lorentz violation.

\begin{figure}[t]
\begin{center}
\psfig{file=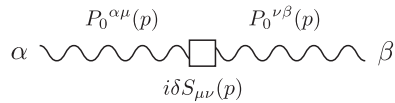,width=2.0in}
\end{center}
\caption{Propagator insertion.
The wavy lines denotes the conventional Lorentz-invariant Stueckelberg propagator $P_0$. 
The square box represents the leading-order Lorentz-breaking insertion 
given by Eq.~(\ref{delta-S}).
}
\label{fig}
\end{figure}

The exact expression for the corresponding propagator 
is somewhat unwieldy.\cite{Cam12} 
But 
for most applications only leading-order Lorentz-violating effects need to be taken into account.
The Lorentz-breaking contributions can then be incorporated via
this propagator insertion (see also Fig.~\ref{fig}):\cite{Cam12}
\begin{eqnarray}
\label{delta-S}
\delta S(p)^{\mu\nu} & = & 
2i(k_{AF})_\alpha\epsilon^{\alpha\beta\mu\nu}p_\beta
-2(k_{F})^{\alpha\mu\beta\nu}p_\alpha p_\beta\nonumber\\
&&{}+m^2\hat{G}^{\mu\nu}
-\frac{1}{\xi}\big(\hat{G}^\mu_\alpha\, p^\alpha p^{\nu}+\hat{G}^\nu_\alpha\, p^\alpha p^{\mu}\big)\,.
\end{eqnarray}

\section*{Acknowledgments}
This work has been supported in part 
by the Indiana University Center for Spacetime Symmetries, 
by Universidad Andr\'es Bello under Grant No.~UNAB DI-27-11/R, 
as well as by the Portuguese Funda\c c\~ao para a Ci\^encia e a Tecnologia.

\end{document}